# Energy Spectra of Magnetostatic Oscillations in Ferrite Disk Resonators

E.O. Kamenetskii, R. Shavit, and M. Sigalov

Department of Electrical and Computer Engineering,

Ben Gurion University of the Negev, Beer Sheva, 84105, ISRAEL

Ferromagnetic resonators with short-wavelength, so-called magnetostatic (MS), oscillations can be considered in microwaves as point (with respect to the external electromagnetic fields) particles. It was shown recently [E. O. Kamenetskii, Phys. Rev. E, **63**, 066612 (2001)] that MS oscillations in a small ferrite disk resonator can be characterized by a *discrete spectrum of energy levels*. This fact allows analyzing the MS oscillations similarly to quantum mechanical problems. In this paper we give the results of energy spectrum calculations for MS oscillations in a ferrite disk resonator.

# 1. INTRODUCTION

A vast range of problems in condensed medium electrodynamics is due to different properties of atoms and material structures in natural media. It is highly desirable to have new materials that exhibit novel electromagnetic properties and, therefore, may give a basis for new, unexpected applications. Such novel electromagnetic properties, unknown for natural media, can be found in artificial systems and should arise from new properties of particles (atoms) that compose a material structure.

Ferromagnetic resonators with short-wavelength, so-called magnetostatic (MS), oscillations [1] can be considered in microwaves as point (with respect to the external electromagnetic fields) particles. Recently, one of the authors put forth an idea that small ferromagnetic MS resonators with special-form surface metallizations can be considered in microwaves as point particles with properties of local internal magnetoelectric (ME) coupling [2,3]. Such artificial ME particles [that one can consider as glued pairs of small (quasi-static) electric and magnetic dipoles] do not exist in nature. As we have shown in [4-6], composite materials based on these ME particles may exhibit new (unknown for natural condensed media) electromagnetic properties (such as balance of energy, effect of nonreciprocity, symmetry properties of the fields, etc.). Experimental investigations carried out recently have shown that quasistatic microwave ME effect really exists in point particles based on ferromagnetic resonators with special-form surface metallizations, as it has been theoretically predicted. These artificial atoms are characterized by rich spectrums of ME oscillations that can be excited by external RF electric and magnetic fields and their combinations [7-10].

The observed spectrums of ME oscillations in ferrite resonators with surface electrodes arise from the MS oscillation spectrums in "pure" (without surface electrodes) ferrite resonators. So, an analysis of MS spectrums becomes a decisive factor in these investigations. Rich multi-resonance spectrums of magnetostatic (MS) oscillations in small ferrite disk resonators excited by the microwave magnetic field, were experimentally observed more than 40 years ago [11]. Since then, the effect of conversion of electromagnetic power into MS-wave power spectrums (with two-four order differences in the wavelength) in non-spherical ferrite samples was a topic of serious experimental and theoretical investigations [12-14]. In these works the main aim was to show that the multi-resonance MS oscillations could chiefly be observed due to the nonuniform internal DC magnetic field in disk-shaped resonators. The role of the nonuniform internal DC magnetic field as the *principal factor* for the observing rich multi-resonance spectrums has to be subjected, however, to a serious criticism. Recently, we have shown that MS oscillations in a small ferrite disk resonator can be characterized by a *discrete spectrum of energy eigenstates* [15]. This fact reveals absolutely another mechanism of conversion of electromagnetic power into MS-wave power spectrum than it was expressed in [12-14], and allows analyzing the MS oscillations similarly to quantum mechanical problems. It gives a basis for a clearer understanding the nature of the observed multi-resonance spectrum and displays, at the same time, a very important aspect of the artificial electromagnetic material principles: The microscopic properties of artificial electromagnetic materials should be based on the non-electromagnetic (quantum mechanical like) laws. Concerning the problem, one should keep in his mind the fact that when in classical electrodynamics structures the spectral problems are characterized by *wavenumbers* or *squared frequencies* as spectral parameters, in quantum mechanical structures there are *energy eigenstates* as spectral parameters. One also has to take into account that the structural properties of natural materials are determined by total energies of electronic systems.

Based on the so-called model of an "open ferrite disk resonator" (the OFDR model) we have shown in [15] that the nature of the observed MS multi-resonance spectrum originates from the energy-eigenstate oscillations. In scope of this model we are unable, however, to take into account the role of the DC demagnetizing field variation. At the same time, in [14] the authors calculated *positions* of absorption peaks in an open structure with a *non-uniform* (due to the demagnetization effect) DC magnetic field and showed a very good correlation with experimental data. Their analysis of the standing wave conditions in the diametral and the circumferential direction (made without a real explanation of the nature of the observed MS multi-resonance spectrum) resembles the Wilson-Sommerfeld rules of quantization used in the "old quantum theory"[16]. It is well known that in spite of the fact that this "old theory" showed an agreement between the experimentally observed and calculated oscillation peaks, the "new quantum theory" based on the Schrödinger wave equation, gave the deep penetration into the true nature of the quantum rules [16].

In this paper we give the results of energy spectrum calculations for MS oscillations in a ferrite disk resonator. To classify the energy eigenstates we analyze the MS oscillations as a collective-process motion of quasiparticles. Our calculations (made based on the OFDR model) show a good qualitative correlation between the pictures of the calculated and experimental resonance spectrums. The divergence of the quantitative character is supposed to be due to the nonhomogeneity of the internal DC magnetic field. The role of the nonuniform DC magnetic field should be considered as an *additional factor* that, certainly, can lead to distortion of an initial discrete spectrum in a ferrite disk but does not imply a fundamental character.

## 2. RESONANCE FREQUENCIES OF A FERRITE DISK RESONATOR

A model of a normally magnetized open ferrite resonator is shown in Fig. 1. This is a ferrite disk without any perfect electric or perfect magnetic walls. Since the disk has a small thickness/diameter ratio, separation of variables is possible. In a case of such an assumption, we exclude, in fact, an influence of the edge regions.

In a ferrite-disk resonator with a small thickness to diameter ratio, the monochromatic MS-wave potential function $\psi$ is represented as [15]:

$$\psi = \sum_{p,q} A_{pq} \widetilde{\xi}_{pq}(z) \widetilde{\varphi}_q(\rho,\alpha), \qquad (1)$$

where $A_{pq}$ is a MS mode amplitude, $\widetilde{\xi}_{pq}(z)$ and $\widetilde{\varphi}_q(\rho,\alpha)$ are dimensionless functions describing, respectively, "thickness" ($z$ coordinate) and "in-plane", or "flat" (radial $\rho$ and azimuth $\alpha$ coordinates) MS modes. For a certain-type "thickness mode" (in other words, for a given quantity $p$), every "flat mode" is characterized by its own function $\widetilde{\xi}_q(z)$.

Because of separation of variables, one can impose independently the electrodynamical boundary conditions – the continuity conditions for the MS potential $\psi$ and for the normal components of the magnetic flux density – on a lateral cylindrical surface ($\rho = R$, $0 \leq z \leq h$) and plane surfaces ($z = 0$, $z = h$). As a result, we have to solve a system of the following two equations [15]:

$$\tan\left(\beta^{(F)} h\right) = -\frac{2\sqrt{-\mu}}{1+\mu} \qquad (2)$$

and

$$(-\mu)^{\frac{1}{2}} \frac{J'_\nu}{J_\nu} + \frac{K'_\nu}{K_\nu} - \frac{\mu_a \nu}{|\beta^{(F)}| R} = 0. \qquad (3)$$

Here $\mu$ and $\mu_a$ are, respectively, the diagonal and off-diagonal components of the permeability tensor, $\beta^{(F)}$ is the wave number of a MS wave propagating in a ferrite along the bias magnetic field, $J_\nu, J'_\nu, K_\nu,$ and $K'_\nu$ are the values of the Bessel functions and their derivatives on a lateral cylindrical surface ($\rho = R$, $0 \leq z \leq h$).

Equations (2) and (3) correspond, respectively, to characteristic equations for MS waves in a normally magnetized ferrite slab [17] and in an axially magnetized ferrite rod [18]. To obtain eigen frequencies of a ferrite disk resonator one has to solve a system of two equations, Eqs. (2) and (3), for given values of $h$, $R$, and $\nu$. The solutions for oscillating MS modes take place only for $\mu < 0$. It means that the admissible frequency region is restricted by frequencies $\omega_1$ and $\omega_2$ ($\omega_1 \leq \omega \leq \omega_2$), where $\omega_1 = \gamma H_i$ and $\omega_2 = \gamma \left[ H_i (H_i + 4\pi M_s) \right]^{1/2}$. Here $\gamma$ is the gyromagnetic ratio, $H_i$ is the internal DC magnetic field, and $M_s$ is the saturation magnetization. It becomes clear [see Eq. (3)] that one should have *different resonances for the left-hand and the right-hand circularly polarized oscillations* (having different signs of $\nu$).

Fig. 2 (a) illustrates the graphical solutions of Eqns. (2) and (3) obtained for a set of "thickness modes" ($p$ numbers) and different "in-plane (flat) modes" with $\nu = \pm 1$ and with a number of radial variations ($q$ numbers). An analysis was made with use of the disk data given in paper [14]: $4\pi M_s = 1790\, G$, $2a = 3.98\, mm$, $h = 0.284\, mm$. Calculations were made for the external DC magnetic field $H_0 = 5.02\, kOe$. One can see that in our case of a ferrite disk with a small thickness/diameter ratio, the spectrum of "thickness modes" is very "rare" compared to the "dense" spectrum of "flat modes". The entire spectrum of "flat modes" is completely included into the wave-number region of a fundamental "thickness mode". It means that the spectral properties of a resonator can be entirely described based on consideration of only a fundamental "thickness mode". The spectrum of resonance peaks corresponding to solutions of Eqns. (2) and (3) for a fundamental "thickness mode" is shown in Fig. 2 (b) by vertical lines. There is a clear evidence for a strong difference in positions of peaks with positive and negative signs of $\nu$. Such the difference of resonances for the left-hand and the right-hand circularly polarized oscillations reveals a serious problem in an analysis of the energy spectra.

## 3. EIGENSTATES OF MS WAVES AND OSCILLATIONS

The state of the MS-wave system can be represented by the MS-potential function. In an axially magnetized ferrite rod and in a normally magnetized ferrite disk, one can consider the MS potential functions as the probability distribution functions describing by the Schrödinger-like equation. In this case, one obtains the normalized spectrum of energy eigenstates[15]. The MS-wave process can be described making use of the language of motion of certain quasiparticles. In such the description, a clear definition of effective masses of these quasiparticles should be given to be able to characterize the energy spectra. To analyze eigenstates of MS oscillations in a ferrite disk, we should start with consideration of the states of MS waves in an axially magnetized ferrite rod. After that we will extend our analysis for a normally magnetized ferrite disk. In our analysis of eigenstates of MS oscillations we should distinguish two cases: (a) the case of a constant-value bias magnetic field with a frequency variation and (b) the case of a constant-value frequency with a bias magnetic field variation.

### (A) An axially magnetized ferrite rod

By appropriate change of variables, any system of equations describing oscillations in *one-dimensional linear structures with distributed parameters* may be written as (see, for example, [19]):

$$\hat{Q}\vec{u} = \frac{\partial \vec{u}}{\partial t}, \qquad (4)$$

where $\vec{u}(z,t)$ is a vector function with components $u_1, u_2, \ldots$ describing the system properties and $\hat{Q} = \hat{Q}(z)$ is a differential-matrix operator. In a general case, the $jk$ matrix element of operator $\hat{Q}(z)$ is written as:

$$Q_{jk}(z) = a_{jk}^{(m)}(z)\frac{\partial^m}{\partial z^m} + a_{jk}^{(m-1)}(z)\frac{\partial^{m-1}}{\partial z^{m-1}} + \ldots + a_{jk}^{(1)}(z)\frac{\partial}{\partial z} + a_{jk}^{(0)}(z), \qquad (5)$$

where *m* is an order of every differential equation of a system (4). A number of boundary conditions necessary to solve a system is a total order of equations (4). In a case of an electromagnetic process of the free-space plane-wave propagation, for example, Eqns. (4) and (5) correspond to Maxwell's equations written for transversal components of $\vec{E}$ and $\vec{H}$ - fields.

Suppose that oscillations in a one-dimensional linear structure are described by scalar wave function $\psi$. In a case of a lossless structure, one has from Eqns. (4) and (5):

$$a^{(1)}(z)\frac{\partial^2 \psi(z,t)}{\partial z^2} + a^{(2)}(z)\psi(z,t) = \frac{\partial \psi(z,t)}{\partial t} \qquad (6)$$

This is the Schrödinger-like equation. The solution of this equation can be found as a product of functions dependent only on *z* and only on *t*. Since the left-hand side of Eqn. (6) is the function dependent only on *z* and the right-hand side – only on *t*, one can conclude that both these sides should be equal to the same constant value. This makes possible to consider Eqn. (6) as the *stationary-state* equation.

Let a one-dimensional linear structure be a waveguide structure with parameters not dependent on longitudinal *z* coordinate. So coefficients $a^{(1)}$ and $a^{(2)}$ in Eqn. (6) are not dependent on *z*. We consider a MS-wave waveguide based on an axially magnetized ferrite cylinder. The feature of this waveguide structure is the fact that there are two cutoff frequencies $\omega_1, \omega_2$. For a given frequency in the frequency region between the cutoff frequencies ($\omega_1 \leq \omega \leq \omega_2$), one has a *complete discrete spectrum of propagating MS modes* [15,18]. So for a monochromatic process ($\psi \sim e^{i(\omega t - \beta z)}$), the $\psi$ function is expanded by the complete-set membrane functions $\tilde{\varphi}$ of MS-wave waveguide modes. In this case we have an infinite set of differential equations [everyone is similar to Eqn. (6)] written for waveguide modes. For frequency $\omega$ and for a certain *n*-th waveguide mode, we obtain from Eqn. (6):

$$-a_n^{(1)}\beta_n^2 + a_n^{(2)} = i\omega \qquad (7)$$

For harmonic processes, coefficients $a^{(1)}$ and $a^{(2)}$ should be imaginary quantities. There is, however, certain vagueness how to determine these coefficients.

Let us represent a MS-potential function as a quasi-monochromatic quantity:

$$\psi = \psi^{(\max)}(z,t)\, e^{i(\omega t - \beta z)}, \tag{8}$$

where amplitude $\psi^{(\max)}(z,t)$ is a smooth function of longitudinal coordinate and time, so that

$$\left|(\beta^{-1}\nabla)\psi^{(\max)}\right| \ll \psi^{(\max)}, \quad \left|\left(\omega^{-1}\frac{\partial}{\partial t}\right)\psi^{(\max)}\right| \ll \psi^{(\max)}. \tag{9}$$

Let a part of an infinitely long axially-magnetized-ferrite-rod lossless MS waveguide be restricted by two cross sections placed at $z = z_1, z_2$. For the quasi-monochromatic MS wave process, the energy balance equation in a waveguide section:

$$\int_{z_1}^{z_2}\int_S \nabla_\| \cdot \overline{\vec{P}}_\| \, ds\, dz = -\frac{d}{dt}\int_{z_1}^{z_2}\int_S \overline{w}\, ds\, dz \tag{10}$$

can be rewritten as [15]:

$$\frac{1}{4}i\omega\mu_0 \int_{z_1}^{z_2}\int_S (\psi\nabla_\|^2\psi^* - \psi^*\nabla_\|^2\psi)\, ds\, dz = \frac{d}{dt}\int_{z_1}^{z_2}\int_S \overline{w}\, ds\, dz \tag{11}$$

In the above equations, $\overline{\vec{P}}_\|$ is the average (on the RF period) power for flow density along a MS waveguide, $\nabla_\|$ means the longitudinal part of divergence, and $\overline{w}$ is the average density of energy. Since coefficients $a^{(1)}$ and $a^{(2)}$ are imaginary quantities, one obtains from Eqns. (6) and (11):

$$-\frac{1}{4a^{(1)}}i\omega\mu_0 \int_{z_1}^{z_2}\int_S \left(\psi\frac{\partial\psi^*}{\partial t} + \psi^*\frac{\partial\psi}{\partial t}\right) ds\, dz = \frac{d}{dt}\int_{z_1}^{z_2}\int_S \overline{w}\, ds\, dz \tag{12}$$

For mode $n$, the average energy of MS waveguide section can be characterized as

$$\overline{W}_n = -\frac{1}{4a_n^{(1)}}i\omega\mu_0 \int_{z_1}^{z_2}\int_S \psi_n\psi_n^*\, ds\, dz + C \tag{13}$$

where $C$ is an arbitrary quantity not dependent on time. We can normalize the process in a supposition that constant $C$ is equal to zero.

One can see that coefficient $a_n^{(2)}$ is not included in the expression of average energy. The only coefficient included in this expression is coefficient $a_n^{(1)}$. Another important conclusion following from Eqn. (13) is that for any coefficient $a_n^{(1)}$ the energy can be orthogonolized with respect to the known $\psi_n$ eigenfunctions.

Let us represent the MS-potential function in a ferrite rod as

$$\psi = A\tilde{\varphi}\, e^{-i\beta z} \qquad (14)$$

where $A$ is a dimensional coefficient and $\tilde{\varphi}$ is a dimensionless membrane function. Since membrane functions of MS modes in an axially magnetized ferrite rod give a complete discrete set of functions (on a waveguide cross section), the dimensionless membrane function $\tilde{\varphi}$ can be written as

$$\tilde{\varphi} = \sum_{n=1}^{\infty} b_n \tilde{\varphi}_n, \qquad (15)$$

where $\tilde{\varphi}_n$ is a membrane function of MS mode and $b_n$ are constants. In a case of a cylindrical ferrite rod, $\tilde{\varphi}_n$ are described by the Bessel functions [18]. It is found [15] that the wave function $\tilde{\varphi}$ is normalized to unity when the coefficients $b_n$ satisfy the relation $\sum_n |b_n|^2 = 1$.

Based on the Walker equation, one has for every MS mode in an axially magnetized rod:

$$\hat{G}_\perp \tilde{\varphi}_n = \beta_n^2 \tilde{\varphi}_n, \qquad (16)$$

where

$$\hat{G}_\perp \equiv \mu \nabla_\perp^2 \qquad (17)$$

One can see that for MS modes propagating in a ferrite rod, operator $\hat{G}_\perp$ is the positive definite operator.

Let us introduce a certain quantity $K_n$ and suppose that instead of orthogonal functions $\widetilde{\varphi}_n$ one has a set of functions

$$\widetilde{u}_n = K_n \widetilde{\varphi}_n. \qquad (18)$$

If the norm of functions $\widetilde{\varphi}_n$ is equal to unit, the norm of function $\widetilde{u}_n$ is equal to $|K_n|^2$. Let us apply operator $\hat{G}_\perp$ to functions $\widetilde{u}_n$. We can write:

$$\hat{G}_\perp \widetilde{u}_n = \beta_n^2 \widetilde{u}_n \qquad (19)$$

Since $|K_n|^2$ are real positive numbers, we also have the positive definiteness when operator $\hat{G}_\perp$ is applied to functions $\widetilde{u}_n$.

The fact that coefficient $a^{(2)}$ is not included in the expression of average energy gives us a possibility to consider different cases based on certain physical models. One can see that when $a^{(2)} \equiv 0$, Eqn. (6) resembles the Schrödinger equation for "free particles". This is the case of a constant value of bias magnetic field $\vec{H}_0$. Certainly, when a ferrite specimen (having saturation magnetization of a ferrite material), is placed into a bias magnetic field, one has a constant "potential energy" of this ferrite sample in the DC magnetic field. For a given constant value of a bias magnetic field $\vec{H}_0$, the spectral properties of a structure are exhibited with respect to frequency $\omega$. When $a_n^{(2)} \equiv 0$, coefficients $a_n^{(1)}$ are found as [see Eqn. (7)]:

$$a_n^{(1)} = -\frac{i\omega}{\beta_n^2}, \qquad (20)$$

We define a notion of the *normalized average MS energy of mode n* as the average (on the RF period) energy of MS waveguide section with unit length and unit characteristic cross section. This energy for a mode with unit amplitude $\left(|b_n|^2 = 1\right)$ is expressed based on Eqn. (13) as:

$$E_n^{(lm)} = \frac{1}{4}g\mu_0\beta_n^2 \tag{21}$$

where $g$ is the unit dimensional coefficient, having the same dimension as a squared amplitude $A^2$ [see Expr. (14)]. The meaning of superscript (*lm*) used in Eqn. (21) will be explained below.

With reference to Eqns. (13) and (18), one can see that in this case:

$$|K_n|^2 = \frac{1}{4}g\mu_0 \tag{22}$$

The MS-potential wave function describes possible eigenstates of a system. From the above analysis, the following question, certainly arises: Can the considered above energy quantization (described by the MS-potential properties) be regarded as a collective effect of quasiparticles? In other words: Can the MS-wave phenomena in a special *macrodomain* be simply reduced to the case of a many-particle correlated system? When dealing with quasiparticles, it is standard to introduce the concept of "effective mass", i.e. a quantity with dimension of mass, characterizing dynamic properties of a quasiparticle. A quasiparticle may behave differently in different conditions, so that "effective masses" proliferate.

On the corpuscular language, an oscillating process in a magnetically ordered body is a collection of magnons [1,20]. The magnons, being the quantum quasiparticles, are characterized as the quantized states with the lack of localization in space. The energy of every quasiparticle is

$$\varepsilon = \hbar\omega, \tag{23}$$

where $\omega$ is a frequency of magnetic oscillations. For short-wavelength magnetic oscillations (when the exchange interaction is dominant and when the role of boundary conditions is negligibly small), the magnon can be considered as a free particle with a certain effective mass. In this case, the correlation between energy and a momentum of a magnon is similar to such a correlation for a free non-relativistic particle. For long-wavelength magnetic oscillations (when the exchange interaction is negligibly small) – the MS (or Walker-type) oscillations – one does not have so simple connection between the magnon energy and momentum. One can suppose, however, that the process of MS-wave propagation is considered as the motion process of certain quantum quasiparticles having quantization of energy and characterizing by certain *effective masses*. We, conventionally, will call these quasiparticles as the *"light magnons" (lm)*. The meaning of this term arises from the fact that effective masses of the "light magnons" should be much less than effective masses of the (real, "heavy") magnons – the quasiparticles existing due to the exchange interaction. In our description of MS oscillations we neglect the exchange interaction and the "magnetic stiffness" should be described based on the "weak" dipole-dipole interaction [1,20]. The states of the "light magnons" are described based on the so-called *transitional eigenfunctions* [21]. For these transitional eigenfunctions energy is proportional to a squared wavenumber [21]. In our case this is a squared wavenumber of a propagating MS mode. For MS mode $n$, the number of "light magnons" in a MS waveguide section is defined from Eqns. (21) and (23) as the ratio: $\dfrac{E_n^{(lm)}}{\varepsilon}$. When we juxtapose Eqn. (6) with the Schrödinger equation for "free particles" ($a^{(2)} \equiv 0$), we get the following expression for an effective mass of a "light magnon":

$$m_{eff}^{(lm)} = \frac{i\hbar}{2a^{(1)}} \qquad (24)$$

Taking into account Eqn. (20) (and supposing that in the Schrödinger equation $\psi \sim e^{i\omega t}$) we obtain from Expr. (24) for mode $n$:

$$\left(m_{eff}^{(lm)}\right)_n = \frac{\hbar}{2}\frac{\beta_n^2}{\omega} \qquad (25)$$

This expression looks very similar to an effective mass of the (real, "heavy") magnon for spin waves with the quadratic character of dispersion [1].

In an infinite-ferrite-rod MS-wave waveguide for given frequency $\omega'$ ($\omega_1 \leq \omega' \leq \omega_2$), one has a flow of quasiparticles with different "effective masses" and different "kinetic energies". For another frequency $\omega'' \neq \omega'$ ($\omega_1 \leq \omega'' \leq \omega_2$) we have a flow of another quasi-particles differing from previous ones by "effective masses" and "kinetic energies". At a certain frequency, the total energy of non-interacting quasiparticles is equal to a sum of energies of separate quasiparticles:

$$E_{tot}^{(lm)} = \sum_n E_n^{(lm)} \qquad (26)$$

**(B) A normally magnetized ferrite disk**

Based on the above consideration of the states of MS waves in an axially magnetized ferrite rod, we extend now our analysis to a case of a normally magnetized ferrite disk. As we discussed above, in a ferrite disk with a small thickness/diameter ratio, the spectrum of "thickness modes" is very "rare" compared to the "dense" spectrum of "flat modes". So, the spectral properties of such a resonator can be entirely described based on consideration of only a fundamental "thickness mode".

For a given quantity $H_0$, in a case of an infinite ferrite rod, we had a complete set of transitionally moving "light magnons". For a given quantity $H_0$, in a case of a ferrite disk resonator, we have a set of "light magnons" having the reflexively-translational character of motion. The "light magnons" "exist" only inside a ferrite. So the reflexively-translational motion of the "light magnons" takes place between the planes $z = 0$ and $z = h$. Since at a certain frequency we have two waves propagating forth and back with respect to z-axis, the average energy will be twice more than the energy expressed by Eqn. (21). One has the following expression for the "light-magnon" average energy of "flat" mode $q$ in a normally magnetized ferrite disk:

$$E_q^{(lm)} = \frac{1}{2} g\mu_0 \left(\beta_q^{(F)}\right)^2 \qquad (27)$$

where $\beta_q^{(F)}$ is a MS-wave propagation constant in a ferrite of mode $q$. Computations of these energy levels will be done below. But before starting these calculations, we should overcome the difficulties related to the difference in positions of peaks with positive and negative signs of $\nu$, mentioned above in Section 2 of the paper. The problem can be solved based on the so-called *essential boundary conditions*.

## 4. NATURAL AND ESSENTIAL BOUNDARY CONDITIONS

With consideration of the MS-wave process as a motion of quasiparticles and definition of effective masses of these quasiparticles one cannot, however, calculate the energy spectra because of an ambiguity arising from differences in positions of peaks with positive and negative signs of $\nu$. Such the difference, shown in Fig. 2, reveals a contradiction in formulation of the orthonormality relations. In the analysis, it should be supposed that function $\widetilde{\varphi}$ is a single valued function for angle $\alpha$ varying from 0 to $2\pi$. This means that one should be able to write:

$$\widetilde{\varphi}(\alpha) = \widetilde{\varphi}(\alpha + 2\pi). \qquad (28)$$

If this condition takes place, $\nu$ is an integer, positive or negative, quantity (including zero) and for functions $\widetilde{\varphi}$ one can write the following normalization condition:

$$\int_0^{2\pi} \widetilde{\varphi}_\nu(\alpha)\, \widetilde{\varphi}^*_{\nu'}(\alpha)\, d\alpha = \delta_{\nu\nu'}, \qquad (29)$$

where $\delta_{\nu\nu'}$ is Kronecker delta.

Our initial supposition about single-valuedness of "flat" functions $\widetilde{\varphi}$ meets, however, a certain contradiction. Taking into account, for example, Expr. (21) for average energy $E$ in a ferrite rod, we can describe the radial part of function $\widetilde{\varphi}$ by the following two second-order differential equations:

$$\frac{\partial^2 \widetilde{\varphi}}{\partial \rho^2} + \frac{1}{\rho}\frac{\partial \widetilde{\varphi}}{\partial \rho} + \left(\frac{4E}{g\mu_0\mu} - \frac{\nu^2}{\rho^2}\right)\widetilde{\varphi} = 0 \qquad (30)$$

inside a ferrite region ($\rho \leq R$, where $R$ is a disk radius) and

$$\frac{\partial^2 \tilde{\varphi}}{\partial \rho^2} + \frac{1}{\rho}\frac{\partial \tilde{\varphi}}{\partial \rho} + \left(\frac{4E}{g\mu_0} - \frac{\nu^2}{\rho^2}\right)\tilde{\varphi} = 0 \qquad (31)$$

outside a ferrite region ($\rho \geq R$). The acceptable solutions of these equations (which are regular at $\rho = 0$ and vanish at $\rho = \infty$) are described by piecewise continuous Bessel functions [18]. The homogeneous electrodynamics boundary conditions at $\rho = R$ demand continuity for $\tilde{\varphi}$ and continuity for the radial component of the magnetic flux density. The last boundary condition is described as

$$\mu(H_\rho)_{\rho=R^-} - (H_\rho)_{\rho=R^+} = -i\mu_a(H_\alpha)_{\rho=R^-}, \qquad (32)$$

where $H_\rho$ and $H_\alpha$ are, respectively, radial and azimuth components of the RF magnetic field, $\mu_a$ is the off-diagonal component of the permeability tensor, $R^-$ and $R^+$ correspond, respectively, to ferrite ($\rho \leq R$) and dielectric ($\rho \geq R$) regions near the boundary. With use of the magnetostatic solutions: $H_\rho = -\frac{\partial \psi}{\partial \rho}$ and $H_\alpha = -\frac{1}{\rho}\frac{\partial \psi}{\partial \alpha}$, one can rewrite (32) as

$$\mu\left(\frac{\partial \tilde{\varphi}}{\partial \rho}\right)_{\rho=R^-} - \left(\frac{\partial \tilde{\varphi}}{\partial \rho}\right)_{\rho=R^+} = -\frac{\mu_a}{R}\nu(\tilde{\varphi})_{\rho=R^-} \qquad (33)$$

This is a special boundary condition on the border "in-plane" contour $L$. Really, the "flat" functions $\tilde{\varphi}$ determined by two second-order differential equations (30) and (31) should be degenerated with respect to a sign of $\nu$. At the same time, in accordance with a first-order differential equation (33), the functions $\tilde{\varphi}$ are *dependent on a sign of $\nu$*. So (because of the boundary conditions) we have different functions $\tilde{\varphi}$ for positive and negative directions of an angle coordinate when $0 \leq \alpha \leq 2\pi$. In other words, for a given sign of $\mu_a$ one can distinguish the *"right"* and the *"left" functions* $\tilde{\varphi}$. It means that functions $\tilde{\varphi}$ cannot be considered as *single-valued functions*. However, following axioms of quantum mechanics [21], each state function, as well as a superposition of the state functions *must be a single-valued analytic expression* satisfying the boundary conditions for the given system. The fact that solution of our problem is dependent on both a modulus and a sign of $\nu$ rises a question about validity of energy orthonormality relation for functions $\widetilde{\varphi}$. Our calculations show different resonance frequencies for the left-hand and right-hand circularly polarized MS oscillations in a ferrite disk. So one can suppose that for modes with different circulation directions the *degenerated energetic spectrums* exist.

In a paper [15], the energy orthonormality relations were obtained as a combined consideration of two kinds of boundary problems: (a) based on a differential-matrix operator with two first-order differential equations and (b) based on one second-order differential equation. In both cases, the homogeneous electromagnetic boundary conditions (a continuity of MS potential and a normal component of magnetic flux density) are used. It becomes evident, however, that the energy orthonormality relations can be obtained based on the another-type boundary conditions. Really, since a two-dimensional ("in-plane") differential operator $\hat{G}_\perp$ contains $\nabla_\perp^2$ (the two-dimensional, "in-plane", Laplace operator), a double integration by parts (the Green theorem) on $S$ – a square of an "in-plane" cross section of an open ferrite disk – of the integral $\int (\hat{G}_\perp \widetilde{\varphi}) \widetilde{\varphi}^* \, dS$, gives the following boundary condition for the energy orthonormality:

$$\mu \left( \frac{\partial \widetilde{\varphi}}{\partial \rho} \right)_{\rho=R^-} - \left( \frac{\partial \widetilde{\varphi}}{\partial \rho} \right)_{\rho=R^+} = 0 \qquad (34)$$

or

$$\mu (H_\rho)_{\rho=R^-} - (H_\rho)_{\rho=R^+} = 0. \qquad (35)$$

For operator $\hat{G}_\perp$, the boundary condition of the MS-potential continuity together with boundary condition (34) [or (35)] is the so-called *essential* boundary conditions [22]. When such boundary conditions are used, the MS-potential eigen functions of operator $\hat{G}_\perp$ form a *complete basis in an energy* *functional space*, and the functional describing an average quantity of energy, has a minimum at the energy eigenfunctions [22].

The boundary conditions of the MS-potential continuity together with boundary condition (32) [or (33)] – the boundary conditions from the domain of definition of operator $\hat{G}_\perp$ – are the so-called natural boundary conditions [22].

We will calculate now the resonance peak positions in a ferrite disk resonator based on the essential boundary conditions, considered above. In this case of boundary conditions one does not have the difference of resonances for the left-hand and the right-hand circularly polarized oscillations.

The resonance peak positions should be obtained based on the graphical solution of Eqn. (2) and the modified-form equation (3). The last one, taking into account the essential boundary conditions, is represented as

$$(-\mu)^{\frac{1}{2}} \frac{J'_\nu}{J_\nu} + \frac{K'_\nu}{K_\nu} = 0 \tag{36}$$

The feature of this equation is the fact that in the point where $\mu = -1$, one has an identity.

Fig. 3 (a) illustrates the graphical solutions of Eqns. (2) and (36) obtained for the main "thickness mode" and different "in-plane (flat) modes" calculated for Bessel functions of order $\nu = 1$ and with a number of radial variations ($q$ numbers). An analysis was made with use of the same data as for calculations shown in Fig. 2. At the frequency corresponding to the quantity $\mu = -1$ one has a break. The spectrum of resonance peaks corresponding to solutions of Eqns. (2) and (36) for a fundamental "thickness mode" is shown in Fig. 3 (b) by vertical lines.

Usually, in experiments the spectral properties of small ferrite resonators are exhibited with respect to a quantity of bias magnetic field, remaining a quantity of frequency without any variations (see, for example, [11,14]). The graphical solutions of Eqns. (2) and (36) obtained for the same data of disk parameters as calculations in Fig. 3, but with respect to the internal DC magnetic field $H_i$, are shown in Fig. 4 (a). The working frequency is . The break takes place at the magnetic field where $\mu = -1$. The spectrum of resonance peaks is shown in Fig. 4 (b) by long vertical lines. Short vertical lines in Fig. 4 (b) correspond to the Yukawa and Abe's experimental resonance peaks [14]. One can note a good qualitative correspondence of the experimental and calculated spectral pictures. The differences in positions of the first-order (experimental and calculated) peaks are due to the non-homogeneity of the internal DC magnetic field that was not taken into account in the present model.

## 5. CORRELATION BETWEEN THE FREQUENCY AND MAGNETIC-FIELD SPECTRA IN A FERRITE DISK RESONATOR

In a general consideration, the physical justification for definition of the energy levels in a ferrite disk should be based on the notion of the *density matrix* used in quantum mechanics [20, 22]. As we discussed above, the $\psi$ function can be expanded by complete-set membrane functions $\widetilde{\varphi}$ of MS-wave waveguide modes for a monochromatic process ($\psi \sim e^{i\omega t}$). In this case we have an infinite set of differential equations [everyone is similar to Eqn. (6)] written for waveguide modes. This is not the situation one may see in a ferrite disk resonator.

Let us consider the case of a constant-value bias magnetic field $H_0$. Every resonance mode in Fig. 3 is described by Eqn. (6). However, since every resonance peak is characterized by its own frequency, one can suppose that there are no complete-set membrane functions $\widetilde{\varphi}$ of MS modes. For a given quantity of bias magnetic field $H_0$, let us introduce the following function $\Theta(\omega_a, \omega_b)$ written for the "in-plane" functions $\widetilde{\varphi}$:

$$\Theta(\omega_a, \omega_b) = \int_S \widetilde{\varphi}^*(\omega_a) \widetilde{\varphi}(\omega_b) ds, \qquad (37)$$

where $\omega_{a,b}$ are frequencies of some two resonance peaks (the peaks numbered as *a* and *b*) in Fig. 3. By analogy with the quantum mechanics problems, we will call function $\Theta(\omega_a, \omega_b)$ as the density matrix. Evidently, the density matrix is characterized by the Hermitian property:

$$\Theta^*(\omega_a, \omega_b) = \Theta(\omega_a, \omega_b) \tag{38}$$

Diagonal elements of the density matrix are defined as:

$$\Theta(\omega_a, \omega_a) = \int_S |\tilde{\varphi}(\omega_a)|^2 \, ds \tag{39}$$

In accordance with Fig. 3 one can see that the number of a resonance peak corresponds to the number of an "in-plane" function. So instead of Eqn. (37) one should write:

$$\Theta_{m,n}(\omega_m, \omega_n) = \int_S \tilde{\varphi}_m^*(\omega_m) \tilde{\varphi}_n(\omega_n) \, ds \tag{40}$$

There is no foundation to state a priori that for a given quantity of bias magnetic field $H_0$, when frequencies $\omega_m$ and $\omega_n$ are different (see Fig. 3), functions $\tilde{\varphi}_m$ and $\tilde{\varphi}_n$ are mutually orthogonal.

Since solutions of a system of Eqns. (2) and (36) are found or with respect to the frequency (with the constant-value bias field), or with respect to the bias field (with the constant-value frequency), one has a *mutual matching* between the frequency (see Figs. 3) and the magnetic-field (see Figs. 4) spectra. This matching is illustrated in Figs. 5 for first three peaks ($q$=1,2,3). In Figs. 5 (a), (b), and (c) we see the peak positions for the frequency spectrums, respectively, at $H_0|_{q=1}$, $H_0|_{q=2}$, and $H_0|_{q=3}$ (see Fig. 4). In fact, we sequentially place every peak from the frequency spectrum at the same frequency $f'$ by a sequence of the DC magnetic field values.

For the situation shown in Fig. 5 (when we sequentially place every peak from the frequency spectrum *at the same frequency* ($\omega_m = \omega_n = \omega'$) by a sequence of the DC magnetic field values), the functions $\tilde{\varphi}_m$ and $\tilde{\varphi}_n$ are mutually orthogonal. This statement is evident. Really, in accordance with the above consideration, any "in-plane" eigenfunctions are mutually orthogonal for the monochromatic process. This means that for the magnetic-field (see Figs. 4) spectrum of a ferrite disk, the $\psi$ function can be expanded by complete-set "flat" functions $\widetilde{\varphi}$. So for the magnetic-field spectrum, the eigenfunctions corresponding to the energy eigenstates defined by Expr. (27), are mutually orthogonal.

## 6. ENERGY SPECTRA OF A FERRITE DISK RESONATOR

The energy levels for the "light magnons" were calculated based on Eqns. (27). Fig. 6 shows the positions of quantities $E_q^{(lm)}$ – the "light-magnon" normalized energies – corresponding to different "flat modes" (which have different numbers $q$ for the same quantity $\nu = 1$). The energies found from Eqn. (27) can be considered as "kinetic energies". At the same time, the fact that the spectral properties are exhibited with respect to quantities of a bias magnetic field means variations of "potential energy" of a ferrite sample.

The main feature of the magnetic-field spectrum (see Figs. 4) is the fact that high-order peaks correspond to lower quantities of the DC magnetic field. Physically, the situation looks as follows. Let $H_0^{(1)}$ and $H_0^{(2)}$ be, respectively, the upper and lower values of a bias magnetic field corresponding to the borders of a region, where $\mu < 0$ [1]. Suppose we have a bias field $H_0^{(1)}$. When we put a ferrite sample into this field, we supply it with the energy: $4\pi M_0 H_0^{(1)}$. To some extent, this is a pumping-up energy. *Starting from this point,* we can excite the entire spectrum: from the main mode to the high-order modes. As we move from value $H_0^{(1)}$ to value $H_0^{(2)}$, the energy surplus goes over to the high-order-mode excitation.

Let us calculate the total depth of a "potential well". For the working frequency $\frac{\omega}{2\pi} = 9.51\, GHz$ and saturation magnetization $4\pi M_0 = 1792\, Gauss$ – the data of the Yukawa and Abe's experiments [14] – the depth is calculated as:

$$\Delta U = 4\pi M_0 \left(H_0^{(1)} - H_0^{(2)}\right) = 780\, Oersted \times 1792\, Gauss = 1.4 \cdot 10^6 \frac{ergs}{cm^3} = 14 \cdot 10^4 \frac{Joule}{m^3} \quad (41)$$

As a value of a bias magnetic field decreases, the "particle" obtains the *higher levels of negative energy*. The situation is very resembling the increasing a negative energy of the hole in semiconductors when it "moves" from the top of a valence band [20]. In classical theory, negative-energy solutions are rejected because they cannot be reached by a continuous loss of energy. But in quantum theory, a system can jump from one energy level to a discretely lower one; so the negative-energy solutions cannot be rejected, out of hand.

When one continuously varies the quantity of the DC field $H_0$, for a given quantity of $\omega$, one sees a discrete set of absorption peaks. It means that one has the discrete-set levels of potential energy. This is a very crucial fact that the jumps between the potential levels are controlled (are governed) by the discrete transitions between the quantum states of the "light magnons". This situation can be illustrated based on Figs. 7,8,9. The first three levels ($q = 1,2,3$) of negative potential energy are shown in Fig. 7. These levels were calculated as , where the quantities $H_0|_{q=1,2,3}$ were found from the first-three-peak positions in the magnetic-field spectra (see Figs. 4). For every level of potential energy, the corresponding quantities of the "light-magnon" normalized energies are pointed out. The behaviors of eigenfunctions corresponding to the first three levels ($q = 1,2,3$) are shown in Fig. 8, as the MS-potential distributions along z-axis, and in Fig. 9, as the probability distribution functions $\widetilde{\varphi}\widetilde{\varphi}^*$. Fig. 10 illustrates conformity of the potential energy levels with the "ligt-magnon" energy levels for different quantum numbers.

# 7. DISCUSSION

In the above consideration we showed an analysis of the steady-state functions. To analyze *transitions* between the steady-state energy levels one has to solve Eqn. (6) in the time and space domains. This very interesting problem should be a subject for future investigations. Nevertheless, some aspects of the transitional regimes one can discuss in a frame of this paper.

The average energy of a ferrite disk can be expressed by Eqn. (27) only for a given quantity of a bias magnetic field. With variation of a quantity of bias magnetic field, one has variation of "potential energy" of a ferrite sample. In this case, we should take $a^{(2)} \neq 0$ in Eqn. (6). With a clear similarity with the Schrödinger equation, one can see that coefficient $a^{(2)}(z)$ in Eqn. (6) corresponds to the potential-energy function. So, considering Eqn. (6) as an operator equation with respect to a wave function $\psi$, one can conclude that the first term in the left-hand side of Eqn. (6) describes an operator of kinetic energy, while the second term – the potential energy operator. Since coefficient $a_n^{(2)}$ in Eqn. (7) – the potential energy operator – is dependent on neither frequency $\omega$ nor wavenumber $\beta_n$, one has two ways to define coefficient $a_n^{(1)}$: or by means of taking derivatives over $(\beta_n^2)$ in Eqn. (7), or by means of taking derivatives over $\omega$ in Eqn. (7). As a result, one can write:

$$a_n^{(1)} = -i\frac{1}{d(\beta_n^2)/d\omega} = -i\frac{1}{2}\frac{d^2\omega}{d\beta_n^2} \tag{42}$$

The behavior of quasiparticles characterized by such the coefficient $a^{(1)}$ differs from the behavior of the "light magnons". Such the quasiparticles we will conventionally call as the "*quasimagnons*" *(qm)*. Based on Eqn. (13) one obtains the following expression for the normalized average MS energy of "flat mode" $q$, in a case of "quasimagnons":

$$E_q^{(qm)} = -\frac{1}{2a_q^{(1)}}i\omega\mu_0 g = \frac{1}{2}\omega\mu_0 g \frac{d(\beta_q^{(F)})^2}{d\omega} \tag{43}$$

Here we put coefficient $\frac{1}{2}$ [instead of coefficient $\frac{1}{4}$ used in Eqn. (13)] because of two waves propagating forth and back (with respect to *z*-axis) in a resonator. With reference to Eqns. (13) and (18), coefficient $|K_q|^2$ is determined as

$$|K_q|^2 = -\frac{1}{2a_q^{(1)}} i\omega\mu_0 g \frac{1}{\left(\beta_q^{(F)}\right)^2} = \frac{1}{2}\omega\mu_0 g \frac{d\left(\beta_q^{(F)}\right)^2}{d\omega} \frac{1}{\left(\beta_q^{(F)}\right)^2} \qquad (44)$$

Based on definition of the effective mass for quasiparticles in crystal [20], one can introduce the notion of an effective mass of a "quasimagnon". For MS mode *q* the effective mass of a "quasimagnon" is expressed as:

$$\frac{1}{\left(m_{eff}^{(qm)}\right)_q} = \frac{1}{\hbar} \frac{d^2\omega}{d\left(\beta_q^{(F)}\right)^2} \qquad (45)$$

In accordance with the dispersion properties of MS waves in an axially magnetized ferrite rod [18], one can see that an effective mass of a "quasimagnon" is *negative*. Coefficient $|K_q|^2$ is a positive quantity. An average energy $E_q^{(qm)}$ expressed by Eqn. (43) should be a positive quantity as well. This stipulates the conclusion that to describe the negative-mass "quasimagnons" one should use a notion of negative frequency $\omega$. The "negativeness" of frequency $\omega$ is clearly demonstrated by Fig. 5. One can see that a spectrum "moves" in a negative direction of the frequency axis as we pass from the "top value" of the bias magnetic field.

It is possible to show that the levels calculated based on Eqn. (43) do not give a regular spectral picture. The eigenfunctions corresponding to the energy eigenstates defined by Expr. (43) are not mutually orthogonal. The considered above transitional functions demonstrate some interesting physical aspects, but the real analysis, as we discussed above, should be based on the time- and space-domain calculations.

## 8. CONCLUSION

An analysis of MS oscillation spectra shows that small disk-form ferromagnetic resonators can be considered in microwaves as "artificial atomic structures". The Schrödinger-like equation written for MS-potential wave function shows that in a ferrite disk resonator, MS modes can diagonalize the total magnetic energy. One of the main features of the problem is the fact that one has two types of spectra: (a) the spectra obtained for the constant-value bias magnetic field and (b) the spectra obtained for the constant-value frequency.

Analyzing the MS oscillations similarly to the quantum mechanical problems gives a basis for a clearer understanding the nature of the observed multi-resonance spectrums and displays, at the same time, very important aspects of realization of new artificial electromagnetic materials. It is well known that the structural properties of natural materials are determined by total energies of electronic systems. So novel physical properties of electromagnetic composite materials should arise from energy spectra of MS oscillations in a small ferrite disk.

This paper presents the treatment, which describes the MS-wave system in terms of collection excitation (motion) of quasi-particles – the "light magnons". Such consideration is possible because of discrete energy eigenstates resulting from structural confinement in a special case of a normally magnetized ferrite disk. Confinement phenomena affect the dynamic properties of the magnetic system to a large extent. Recent studies of strongly spatially localized spin wave modes made based on a Brilloin light scattering (BLS) show a large variety of new effects in different-geometry (magnetic wires, magnetic dots) structures [24-26]. An understanding of these phenomena is very important for different new applications of laterally patterned magnetic structures.

**REFERENCES**

1. A. Gurevich and G. Melkov, *Magnetic Oscillations and Waves* (CRC, New York, 1996).

2. E.O. Kamenetskii, Microw. Opt. Technol. Lett., **11**(2), 103 (1996).

3. E.O. Kamenetskii, Phys.Rev. E, **57**, 3563 (1998).

4. E.O. Kamenetskii, Phys. Rev. E, **54**, 4359 (1996).

5. E.O. Kamenetskii, IEEE Trans. Antennas Propag. **49**, 361 (2001).

6. E.O. Kamenetskii, Microw. Opt. Technol. Lett., **19**(6), 412 (1998).

7. E.O. Kamenetskii, I. Awai, A.K.Saha, Microw. Opt. Technol. Lett., **24**(1), 56 (2000).

8. E.O. Kamenetskii, A.K. Saha, I. Awai, IEEE Trans. Magn. **36**, 3485 (2000).

9. A.K., Saha, E.O. Kamenetskii, and I. Awai, Phys. Rev. E, **64**, 056611 (2001).

10. A.K., Saha, E.O. Kamenetskii, and I. Awai, J. Phys. D: Appl. Phys., **35**, 2484 (2002) [see also IoP select, October 2002].

11. J.F. Dillon, Jr., J. Appl. Phys., 31, 1605, (1960).

12. J.R. Eshbach, J. Appl. Phys., 34, 1298 (1963).

13. E. Schlömann, J. Appl. Phys., 35, 159 (1964).

14. T. Yukawa and K. Abe, J. Appl. Phys., 45, 3146 (1974).

15. E.O. Kamenetskii, Phys. Rev. E, 63, 066612 (2001).

16. L. Pauling and E.B. Wilson, *Introduction to Quantum Mechanics* (McGraw-Hill, New York, 1935).

17. R. W. Damon and H. Van De Vaart, J. Appl. Phys., **36**, 3453 (1965).

18. R. I. Joseph and Schlömann, J. Appl. Phys., 32, 1001 (1961).

19. P. S. Landa, Auto-Oscillations in Distributed Systems (Nauka, Moscow, 1983) (in Russian).

20. C. Kittel, Introduction to Solid State Physics, 6th ed. (Wiley, New York, 1986).

21. G. H. Duffey, Quantum States and Processes (Prentice Hall, New Jersey, 1992).

22. S. G. Mikhlin, Variational Metods in Mathematical Physics (Mc Millan, New York, 1964).

23. L. D. Landau and E. M. Lifshitz, Quantum Mechanics: Non-Relativistic Theory, 3rd ed. (Pergamon, Oxford, 1977).

24. J. Jorzick, S. O. Demokritov, B. Hillebrands, M. Bailleul, C. Fermon, K. Y. Guslienko, A. N. Slavin, D. V. Berkov, and N. L. Gorn, Phys. Rev. Lett., 88, 047204 (2002).

25. S. O. Demokritov, B. Hillebrands, and A. N. Slavin, Phys. Rep. 348, 441 (2001).

26. J. Jorzick, C. Kramer, S. O. Demokritov, B. Hillebrands, B. Bartenlian, C. Chappert, D. Decanini, F. Rousseaux, E. Cambril, E. Sondergard, M. Bailleul, C. Fermon, A. N. Slavin, J. Appl. Phys., 89, 7091 (2001).

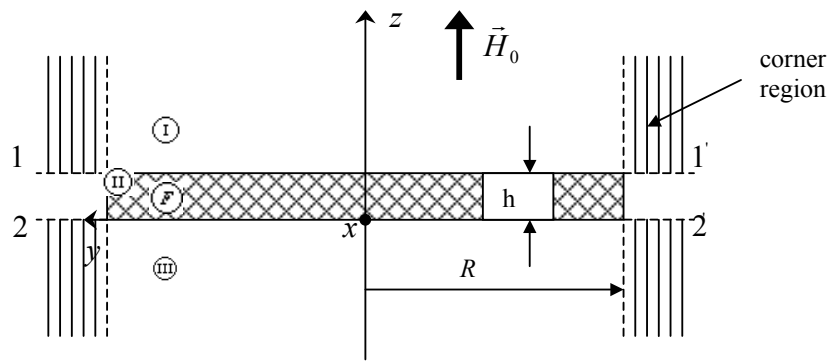

Fig. 1. Normally magnetized ferrite disk with a small thickness/diameter ratio

Paper: "Energy Spectra of Magnetostatic Oscillations in Ferrite Disk Resonators"

Authors: E.O.Kamenetskii, R.Shavit and M.Sigalov

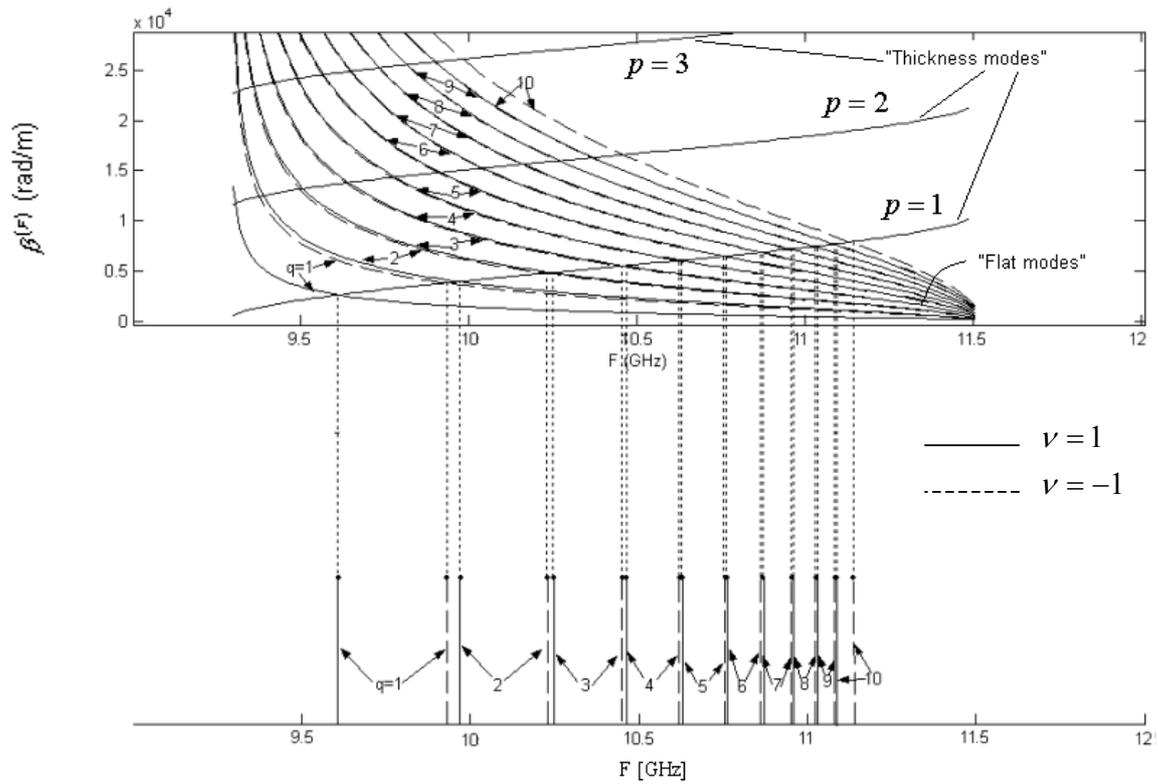

Fig. 2. Resonance frequencies of a ferrite disk resonator
  a) The graphical solution of Eqns. (2) and (3)
  b) The spectrum of resonance peaks for a fundamental
   "thickness mode"

Paper: "Energy Spectra of Magnetostatic Oscillations in Ferrite Disk Resonators"

Authors: E.O.Kamenetskii, R.Shavit and M.Sigalov

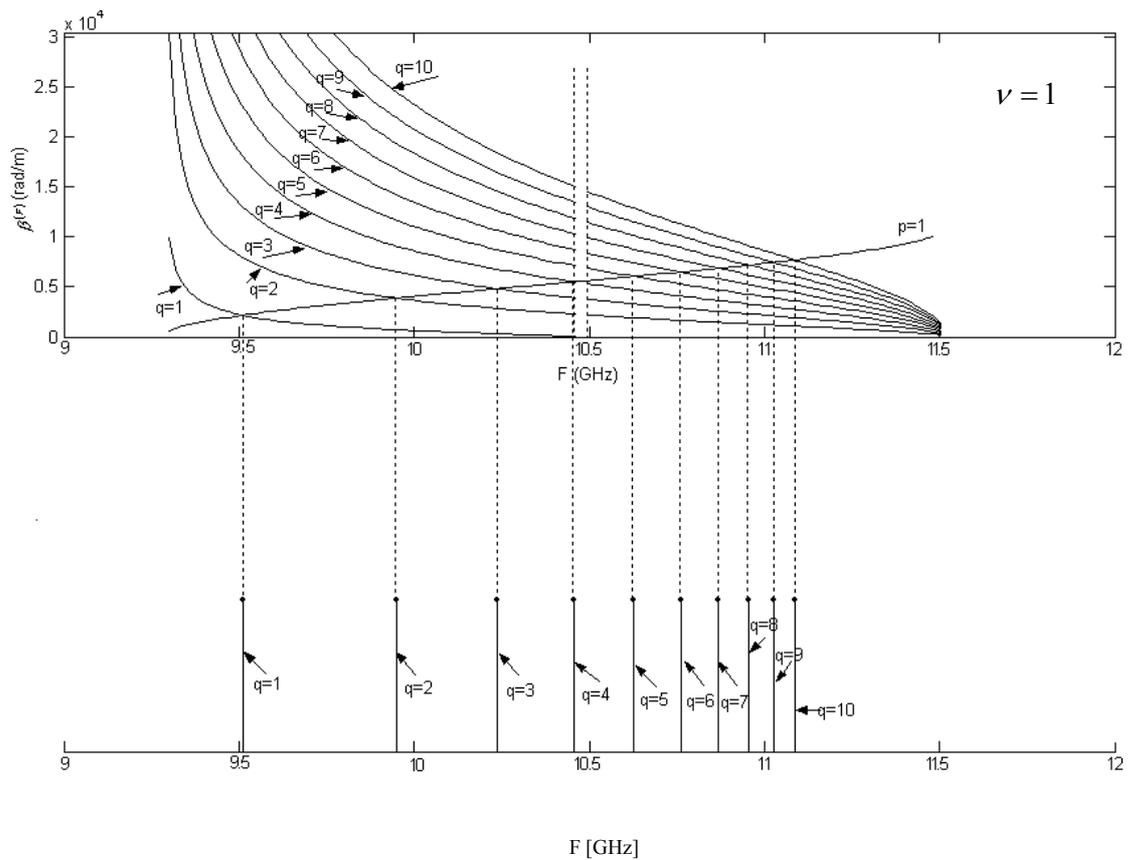

Fig. 3. Resonance spectrum of a ferrite disk vs frequency
(the essential boundary conditions)
a) The graphical solution of Eqns. (2) and (36)
b) The spectrum of resonance peaks for a fundamental
"thickness mode"

Paper: "Energy Spectra of Magnetostatic Oscillations in Ferrite Disk Resonators"

Authors: E.O.Kamenetskii, R.Shavit and M.Sigalov

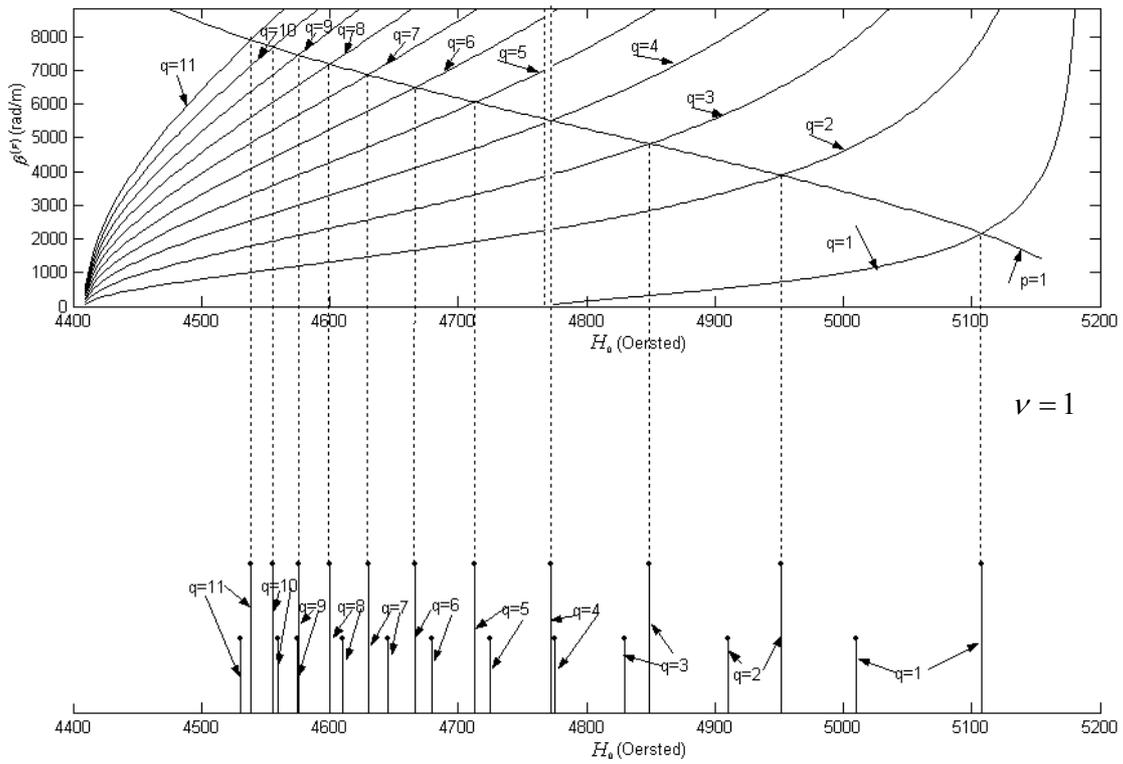

Fig. 4. Resonance spectrum of a ferrite disk vs DC magnetic field
(the essential boundary conditions)
a) The graphical solution of Eqns. (2) and (36)
b) The spectrum of resonance peaks for a fundamental
"thickness mode"

Paper: "Energy Spectra of Magnetostatic Oscillations in Ferrite Disk Resonators"

Authors: E.O.Kamenetskii, R.Shavit and M.Sigalov

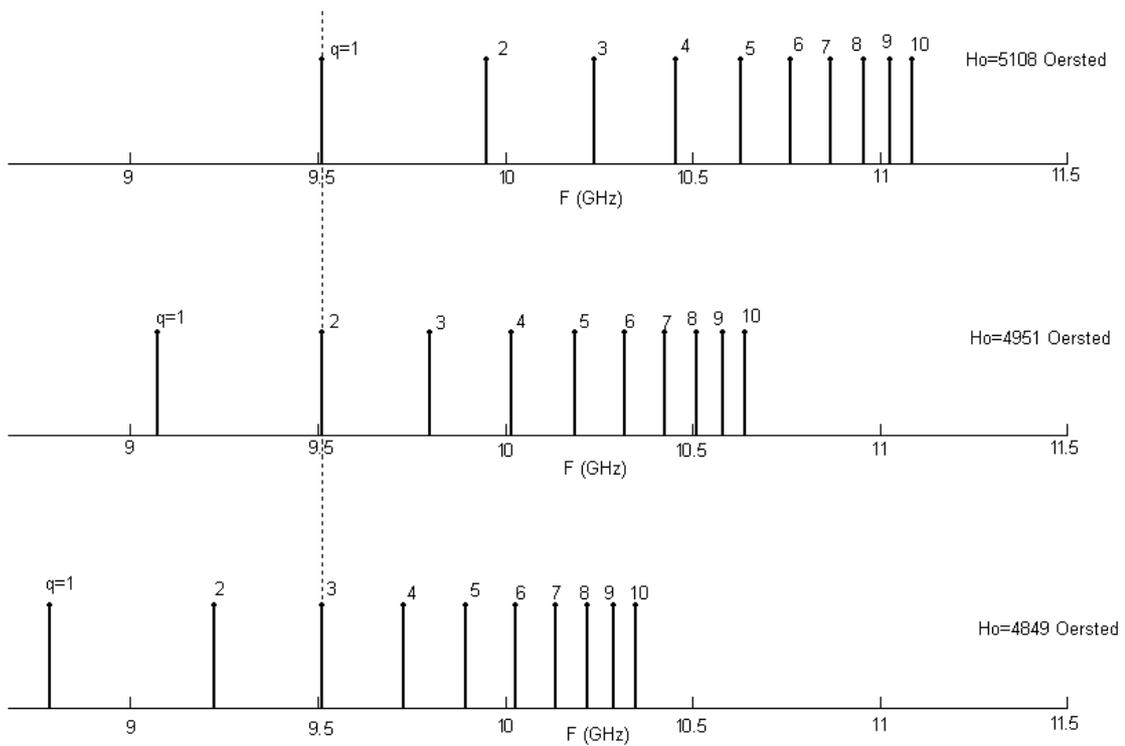

Fig. 5. Mutual matching between the frequency and the magnetic field spectra

Paper: "Energy Spectra of Magnetostatic Oscillations in Ferrite Disk Resonators"

Authors: E.O.Kamenetskii, R.Shavit and M.Sigalov

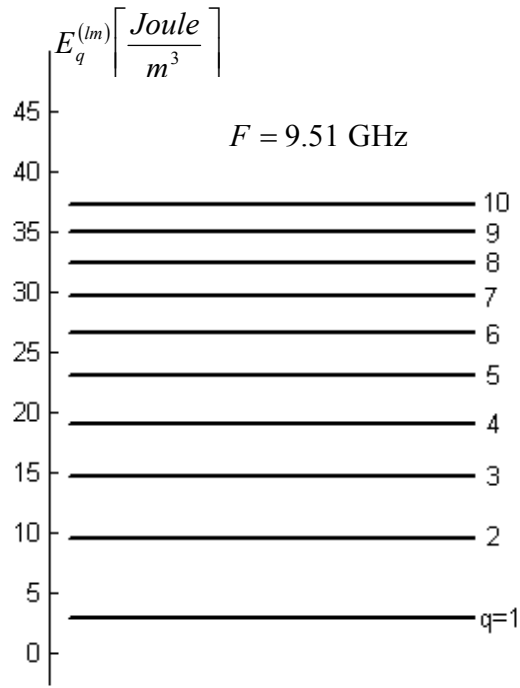

Fig. 6. Levels of the "light-magnon" normalized energies for different modes in a ferrite disk (The case of the constant-value frequency)

Paper: "Energy Spectra of Magnetostatic Oscillations in Ferrite Disk Resonators"

Authors: E.O.Kamenetskii, R.Shavit and M.Sigalov

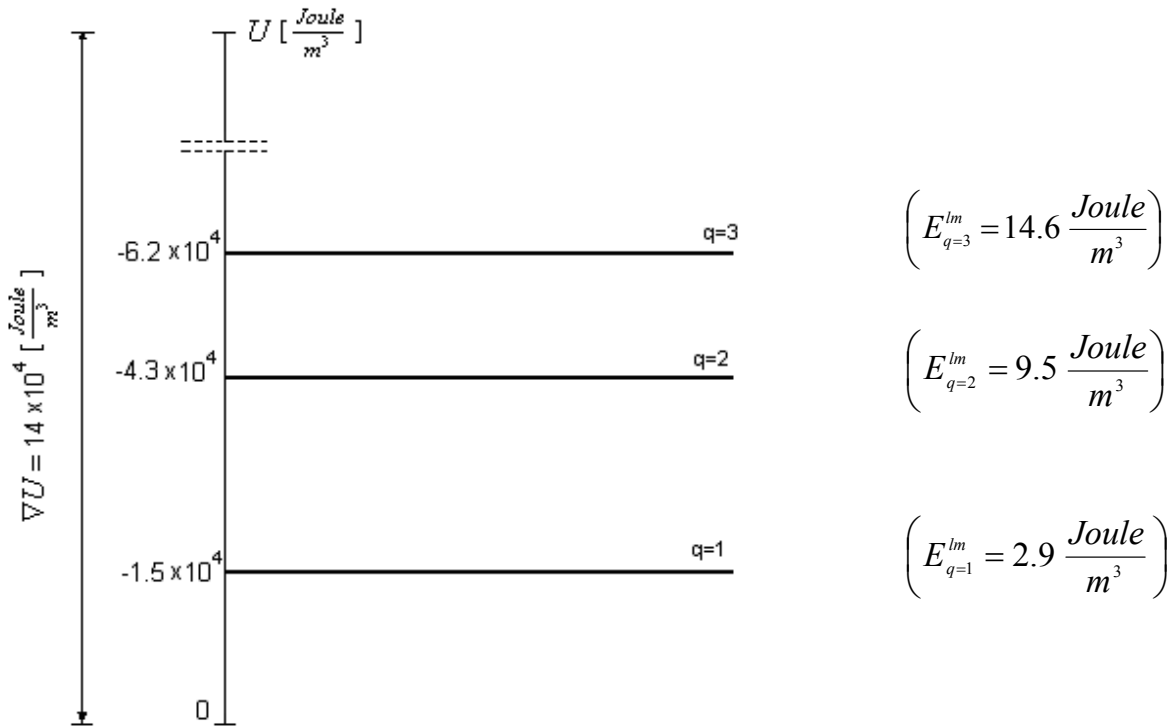

Fig. 7. First three levels of potential energy

Paper: "Energy Spectra of Magnetostatic Oscillations in Ferrite Disk Resonators"

Authors: E.O.Kamenetskii, R.Shavit and M.Sigalov

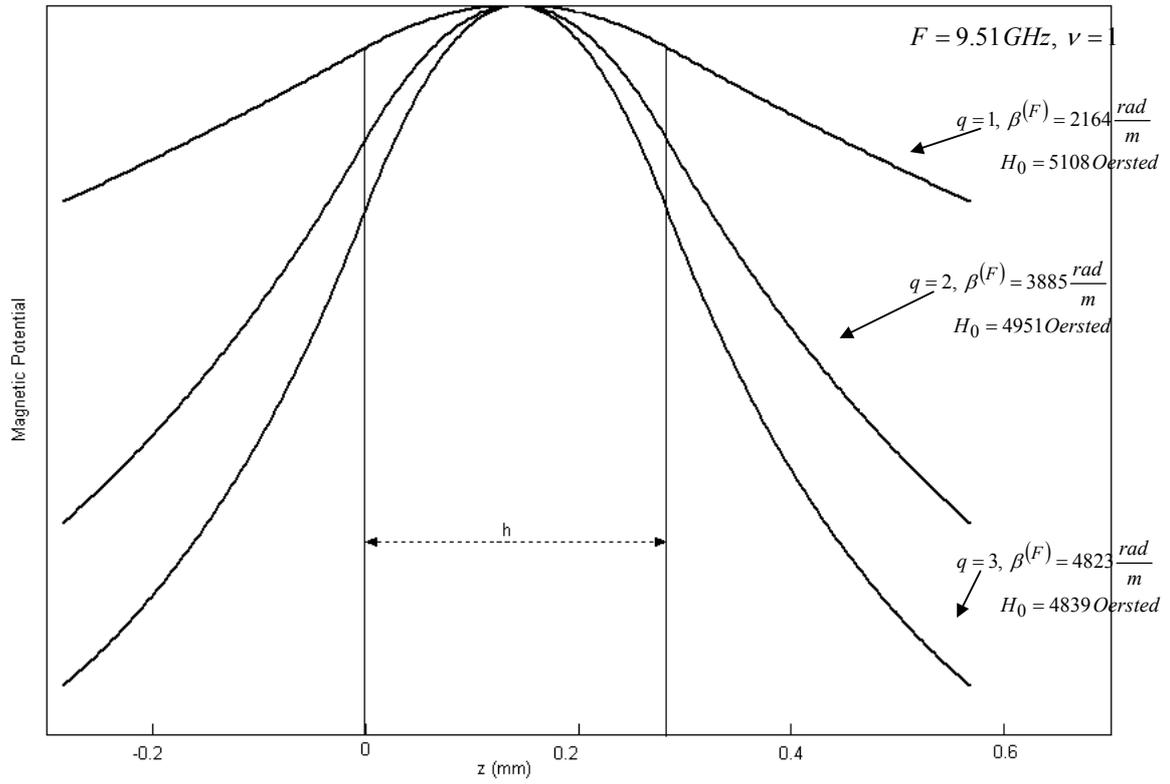

Fig. 8. MS-potential distribution along *z*-axis

Paper: "Energy Spectra of Magnetostatic Oscillations in Ferrite Disk Resonators"

Authors: E.O.Kamenetskii, R.Shavit and M.Sigalov

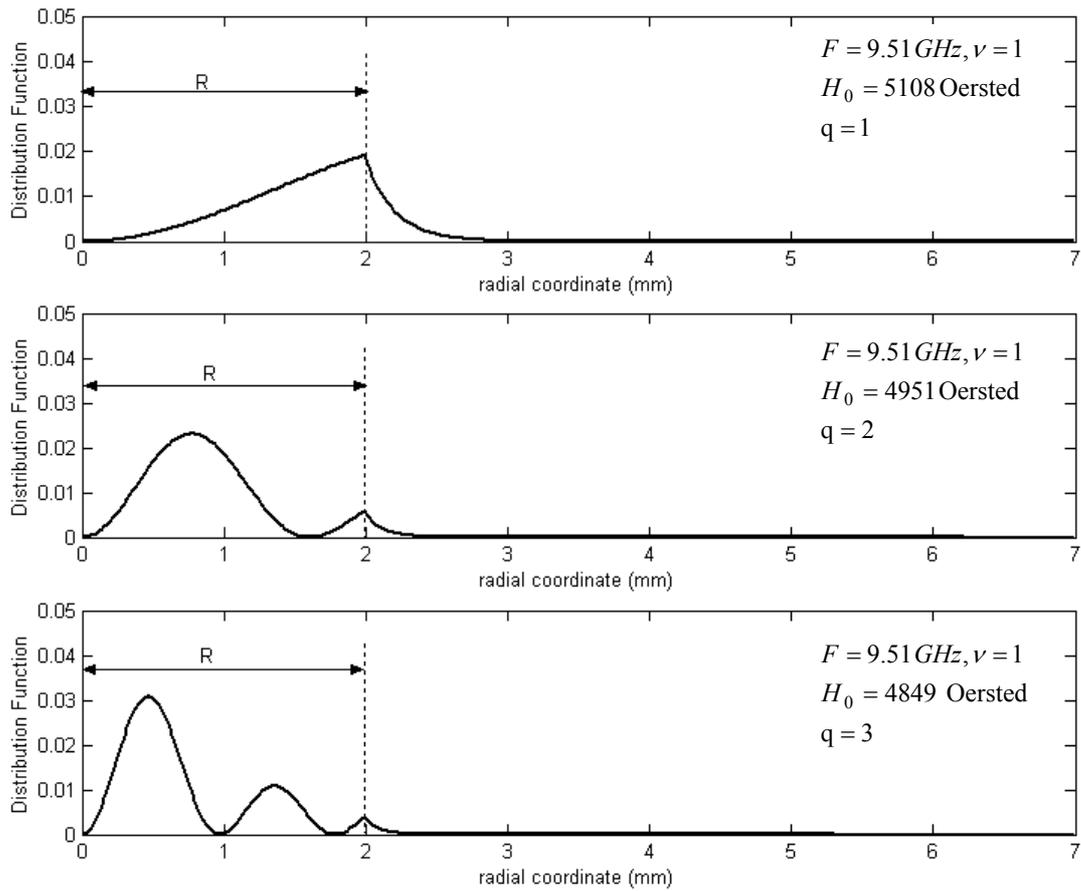

Fig. 9. The probability distribution functions $\widetilde{\varphi}\widetilde{\varphi}^*$ for different "flat modes"

Paper: "Energy Spectra of Magnetostatic Oscillations in Ferrite Disk Resonators"

Authors: E.O.Kamenetskii, R.Shavit and M.Sigalo

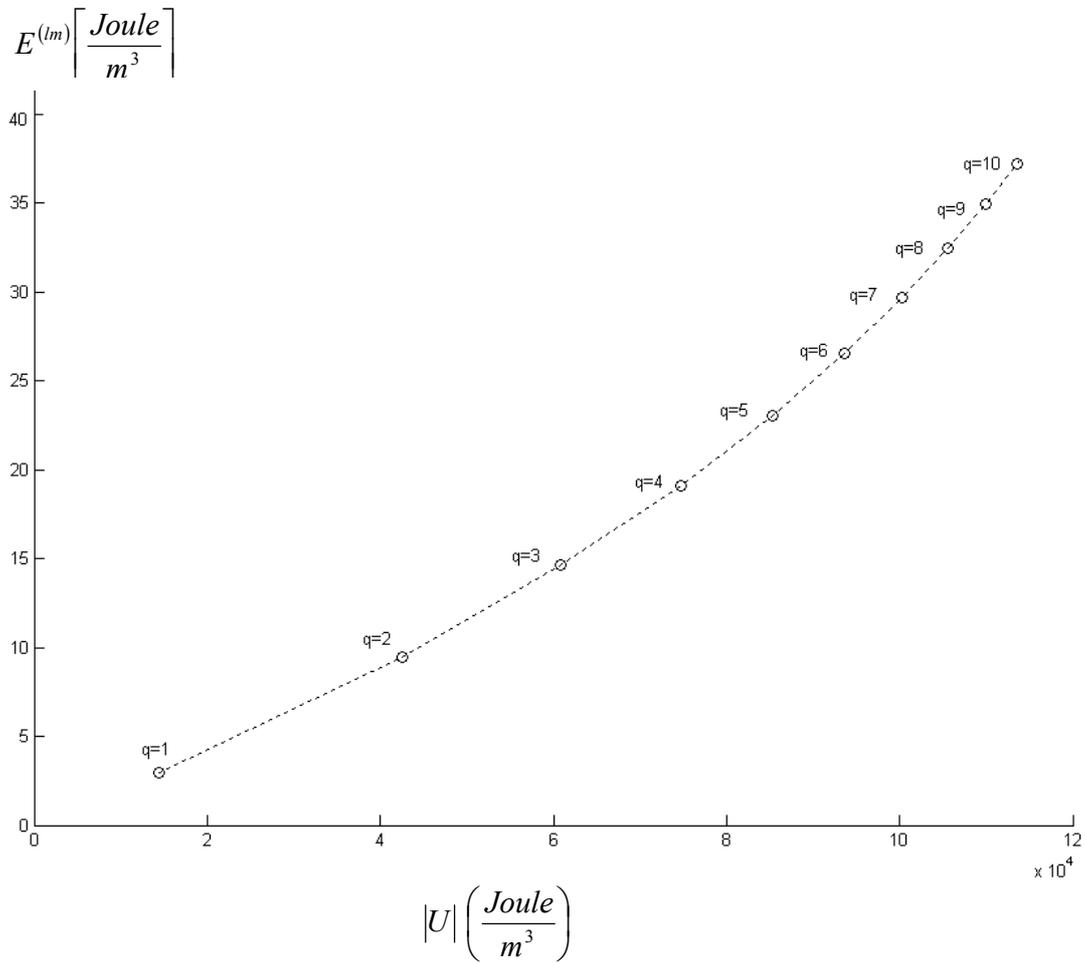

Fig. 10. Conformity of the potential energy levels with the "light-magnon" energy levels

Paper: "Energy Spectra of Magnetostatic Oscillations in Ferrite Disk Resonators"

Authors: E.O.Kamenetskii, R.Shavit and M.Sigalov